\documentclass[10pt]{article}
\usepackage{latexsym,graphicx}
\usepackage{latexsym,graphicx}
\newcommand{\be}{\begin{equation}}
\newcommand{\ee}{\end{equation}}

\newcommand{\Jetset}{\textsc{jetset}}
\def\n{\noindent}
\catcode `\@=11 \catcode `\@=12
\begin{document}
\begin{center}
\large{EXCLUSIVE PION ELECTROPRODUCTION OFF NUCLEONS AND NUCLEI}\\
\vspace{6mm}
\normalsize{MURAT M. KASKULOV and ULRICH MOSEL} \\
\vspace{2mm}
\normalsize{\textit{Institut f\"ur Theoretische Physik, Universit\"at Giessen,
             D-35392 Giessen, Germany \\ 
E-mail : murat.kaskulov@theo.physik.uni-giessen.de}}\\
\end{center}
\vspace{6mm}

\n We present our results for the exclusive electroproduction of pions off
nucleons and nuclei at high values of $Q^2$.
In electroproduction of nuclei $A(e,e'\pi)$ we consider the Color Transparency (CT) effect
recently observed at JLAB. It is shown that the description of $\pi$
production off nucleons is mandatory for the proof of the CT signal at JLAB. 
We further develop the models for exclusive production
of $\pi$ off nucleons at JLAB and HERMES. At first we describe a model based on
exclusive-inclusive connection.  
In the second approach we combine a Regge pole approach with the residual effect of
nucleon resonances.
The nucleon resonances are described using a dual
connection between the exclusive form 
factors and inclusive deep inelastic structure functions. 
We present the results for the beam spin azimuthal asymmetries measured at JLAB
in exclusive electroproduction of charged and neutral pions. 
\vspace{2mm}

\begin{center}
\large{1. \textit{Introduction}}
\end{center}
Electroproduction of mesons in the deep inelastic scattering (DIS),
$\sqrt{s}>2$~GeV and $Q^2>1$~GeV$^2$, 
is a modern tool which permits to study the structure of the nucleon on the partonic
level. Exclusive channels in DIS are of particular importance. 
In this kind of hard processes one may learn about the off-forward 
parton distributions  that parameterize an intrinsic
nonperturbative pattern of the nucleon, see Ref.~\cite{Weiss:2009ar} and references
therein. Much work have been done to 
understand the production of pions in exclusive kinematics.
For instance, in QCD at large values of $(\sqrt{s},Q^2)$
and finite value of Bjorken $x_{\rm B}$  the description of $N(e,e'\pi)N'$ 
relies on the dominance of the longitudinal cross 
section $\sigma_{\rm L}$~\cite{Collins:1996fb}.  The transverse part
$\sigma_{\rm T}$ is
predicted to be  suppressed by power of $\sim 1/Q^2$. 
However, the kinematic domain where this
power suppresion dominates is not
yet known for $\pi$ production. 
A somewhat different concept is used in Regge pole
models which rely on effective hadronic
degrees of freedom. Here
the exclusive $(\gamma^*,\pi)$ forward production mechanism is
is given by the sum of all possible
$t$-channel meson-exchange processes.
Although both partonic and Regge descriptions are presumably dual 
the exclusive reactions
have a potential to discriminate between different models.

Related experimental studies have been carried out at
JLAB~\cite{Horn:2007ug,Collaboration:2010kna} and at HERMES/DESY~\cite{:2007an}.   
A dedicated program on
exclusive production of pions is 
planned in the future at the JLAB upgrade.

On the experimental
side, it is tempting to see an onset of $\sigma_{\rm L}/\sigma_{\rm T} \propto Q^2$ scaling
at presently available energies.
However, the high $Q^2$ data from
JLAB and single spin asymmetries measured in true DIS events at HERMES~\cite{:2009ua}
show nonvanishing
transverse components in $p(\gamma^*,\pi^+)n$.
At JLAB~\cite{Horn:2007ug,:2009ub,Tadevosyan:2007yd,Blok:2008jy}, 
DESY~\cite{Desy,Ackermann:1977rp,Brauel:1979zk}, Cornell~\cite{Cornell_1,Cornell_2,Cornell_3}
and CEA~\cite{{CEA}} the high $Q^2$ region  is dominated by the conversion of
transverse photons in
$\sigma_{\rm T}$.  
For instance, the $Q^2$ dependence of the partial $\sigma_{\rm L}$ and $\sigma_{\rm T}$
cross sections in the
$\pi^{+}$ electroproduction above $\sqrt{s}>2$~GeV  has been studied
in~\cite{Horn:2007ug}. In the
charged pion case  the 
longitudinal cross section $\sigma_{\rm L}$ at forward angles is well
described by the quasi-elastic  $\pi$ knockout
mechanism~\cite{Sullivan:1970yq,Neudatchin:2004pu}. It is driven by
the pion charge form factor~\cite{Horn:2006tm,Huber:2008id} both at JLAB and HERMES.
On the contrary, the $(\sqrt{s},Q^2)$ behavior of $\sigma_{\rm T}$ remains to
be puzzling. The data demonstrate that $\sigma_{\rm T}$ is large and tends to increase relative to
$\sigma_{\rm L}$ as a function of $Q^2$. Interestingly, the $(\sqrt{s},Q^2)$
dependence of exclusive $\sigma_{\rm T}$ exhibits features similar to that in
$p(e,e'\pi^{+})X$ semi-inclusive  
 cross sections in DIS in the limit $z\to
1$~\cite{Kaskulov:2008xc,Kaskulov:2009gp}. This kind of an exclusive-inclusive connection~\cite{Bjorken:1973gc} has been also
observed in exclusive $(\gamma^*,\rho^0)$ production~\cite{Gallmeister:2010wn}.
On the theoretical side, hadronic models based on the meson-exchange scenario
alone largely
underestimate the measured $\sigma_{\rm T}$ in electroproduction, see
Ref.~\cite{Blok:2008jy} for further discussions and references therein. 

We have shown in Refs.~\cite{Kaskulov:2008xc,KM} a possible solution
to the $\sigma_{\rm T}$ problem. The
description of charged pion production proposed in Ref.~\cite{KM} relies on
the residual contribution of the nucleon resonances.  It is supposed that the excitations of nucleon resonances dominate
in electroproduction. The resonances are dual to the direct partonic
interactions due to the Bloom-Gilman duality connection and, correspondingly,
their form factors are determined by parton distribution functions. The $s(u)$-channel resonances
supplement the Reggeon exchanges in the $t$-channel. Therefore, one distinguishes
peripheral $t$-channel meson-exchange processes and the $s(u)$-channel
resonance/partonic 
contributions. In this way all the data collected so far in the
charged pion electoproduction $(e,e'\pi^{\pm})$ at JLAB, DESY, Cornell and CEA
can be well described~\cite{KM}. 

A closely related phenomenon is the color
transparency (CT). It becomes effective in (semi)exclusive electroproduction of mesons
off nuclei, see Ref.~\cite{:2007gqa} for a possible observation and
Refs.~\cite{Larson:2006ge,Cosyn:2007er,Kaskulov:2008ej} for further interpretations of the
CT signal in the reaction $A(e,e'\pi^+)$.

\begin{center}
\large{2. \textit{Color Transparency in semi-exclusive reaction $(e,e'\pi^+)$
off nuclei}}
\end{center}

\begin{figure*}[t]
\begin{center}
\includegraphics[clip=true,width=0.8\columnwidth,angle=0.]
{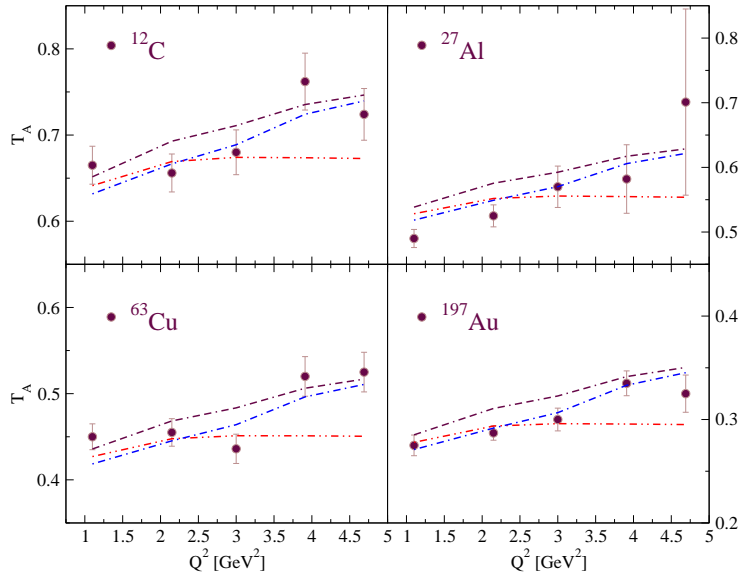}
\caption{\label{Figure8}
\small  Transparency, $T_{A}$, vs. $Q^2$ for $^{12}$C (left, top panel), $^{27}$Al (right, top), $^{63}$Cu (left, bottom) and  $^{197}$Au (right,
bottom). 
The dash--dash--dotted curves realize the CT effect in both the
longitudinal
and transverse channels and dash--dotted curves in the transverse channel only. 
The dot--dot--dashed curves describe the CT effect in the longitudinal
channel only.
The experimental data are from Ref.~\cite{:2007gqa}. 
\vspace{-0.8cm} }
\end{center}
\end{figure*}

The interactions of high--energy virtual photons with nuclei
provide an important tool to study the early stage of hadronization
and (pre)hadronic final--state--interactions (FSI)
at small distances $d \sim 1/\sqrt{Q^2}$.
One believes that at high
values of $Q^2$  the exclusive pions are produced in a point like
configurations which may interact  in the nuclear medium only weakly. 
Thus, in the presence of the CT effect the intranuclear attenuation of hadrons
propagating through the nuclear medium should decrease as a function of photon
virtuality $Q^2$. In this case the nucleus becomes more transparent for the outgoing
particles as compared to the case where the attenuation is driven by
ordinary absorption mechanisms. At JLAB~\cite{:2007gqa} the nuclear
transparency\footnote{The nuclear transparency for a certain reaction process is usually defined
as the ratio of the nuclear cross section per target nucleon to the one for a
free nucleon, i.e. $T_A=\sigma_A/A\sigma_N$.}  in semi--exclusive $\pi^+$ electroproduction reaction
$A(e,e'\pi^+)$ has been measured as a function of $Q^2$ and the atomic mass 
number $A$. A rise of pionic transparency has been indeed observed for values of $Q^2$
between 1 and 5 GeV$^2$, see Fig.~\ref{Figure8}.

In the work of Ref.~\cite{Kaskulov:2008ej} we have studied the onset of CT at JLAB
using a factorization of the whole reaction into an initial, primary
interaction of the incoming virtual photon with the nucleon and the FSI.
Following the model of Ref.~\cite{Kaskulov:2008xc}, the cross section 
for the former is reproduced both in its longitudinal and
its transverse contribution while the FSI is treated within
transport approach. Here
the propagation of the produced (pre)hadron through the nuclear medium is
described by the Boltzmann--Uehling--Uhlenbeck (BUU) equation
which describes the time evolution of the phase space density $f_i(\vec r,\vec
p,t)$ of particles of type $i$ that can interact via binary reactions.
Besides the produced hadron and the nucleons these particles involve the baryonic resonances and mesons
that can be produced in FSI. For the baryons the equation contains a mean
field potential which depends on the particle position and momentum. The BUU
equations of each particle species $i$ are coupled via the mean field and the
collision integral. The latter allows for elastic and inelastic
rescattering and side--feeding through coupled--channel effects; it accounts
for the creation and annihilation of particles of type $i$ in a secondary
collisions as well as elastic scattering from one position in phase space into
another. The resulting system of coupled differential--integral equations
is solved via a test particle ansatz for the phase space density. For fermions
Pauli blocking is taken into account via blocking factors in the collision 
term~\cite{GiBUU}.

We use the quantum diffusion model of Ref.~\cite{farrar} to describe the
time - development of the interactions of a point-like configuration
produced in a hard initial reaction. This approach combines a linear
increase of the hadron--nucleon cross section with the assumption that the
cross section for the leading particles does not start at zero, but at a
finite pedestal value connected with $Q^2$ of the initial interaction, {\it i.e.}
\begin{equation}
\label{effcs}
\sigma^*(t)/\sigma  = X_0
+(1-X_0) \left({(t-t_P)}/{(t_F-t_P)}\right),
\label{eq:scenarioQ}
\end{equation}
where $X_0 = {r_{\rm lead}}\frac{const}{Q^2}$ with $r_{\rm lead}$ standing for
the ratios of leading partons over the total number of partons (2 for mesons,
3 for baryons). The (pre)hadronic expansion times $t_F$ can be extracted from
the string breaking pattern of the Lund model~\cite{Gallmeister:2005ad}. 

In Fig.~\ref{Figure8} we present the results of our calculations. 
The dash--dash--dotted curves realize the CT effect in both the
longitudinal
and transverse channels and dash--dotted curves in the transverse channel only.
In addition we show the results of the CT effect in the longitudinal
channel only (dot--dot--dashed curves). As one can see the latter scenario
is certainly ruled out by the present data. Because of the dominance of the
transverse cross section at high values of $Q^2$, a use of different values
of $t_F$ in a range discussed in~\cite{Kaskulov:2008ej} does not change this result significantly. This is 
particularly interesting because presently the CT effect is expected to show
up in the longitudinal channel~\cite{Strikman:2007nv}.

\begin{center}
\large{3. \textit{Deep exclusive $\pi^+$ electroproduction off nucleons}}
\end{center}

An understanding of $\pi$ production mechanism off nucleons is mandatory 
for the proof of the CT signal observed in the $\pi$
electroproduction off nuclei. As we have seen, the fate of
pions in the nuclear medium depends on the initial longitudinal
and/or transverse production mechanisms~\cite{Kaskulov:2008ej}.

At first we briefly describe the model for the exclusive process
\begin{equation}
\gamma^{*}(q) + N(p) \to \pi(k') + N'(p')
\end{equation}
based on exclusive-inclusive connection.
Following Ref.~\cite{Kaskulov:2008xc} we distinguish two classes of primary
collisions: a soft hadronic and a hard partonic (DIS) production of $\pi^+$.
The soft hadron--exchange part of the $\gamma^* p \to n \pi^+$ amplitude is
described by the exchange of Regge trajectories. However, at the  invariant 
masses reached in the $\pi$CT experiment ($W \approx 2.2$ GeV)  nucleon resonances can contribute to the $1\pi$ channel.
As in Ref.~\cite{Kaskulov:2008xc} this is modeled by the hard interaction
of virtual photons with partons (DIS)  since  DIS involves all possible
transitions of the nucleon from its ground state to any excited state.

\begin{figure}[t]
\begin{center}
\includegraphics[clip=true,width=0.3\columnwidth,angle=0.]
{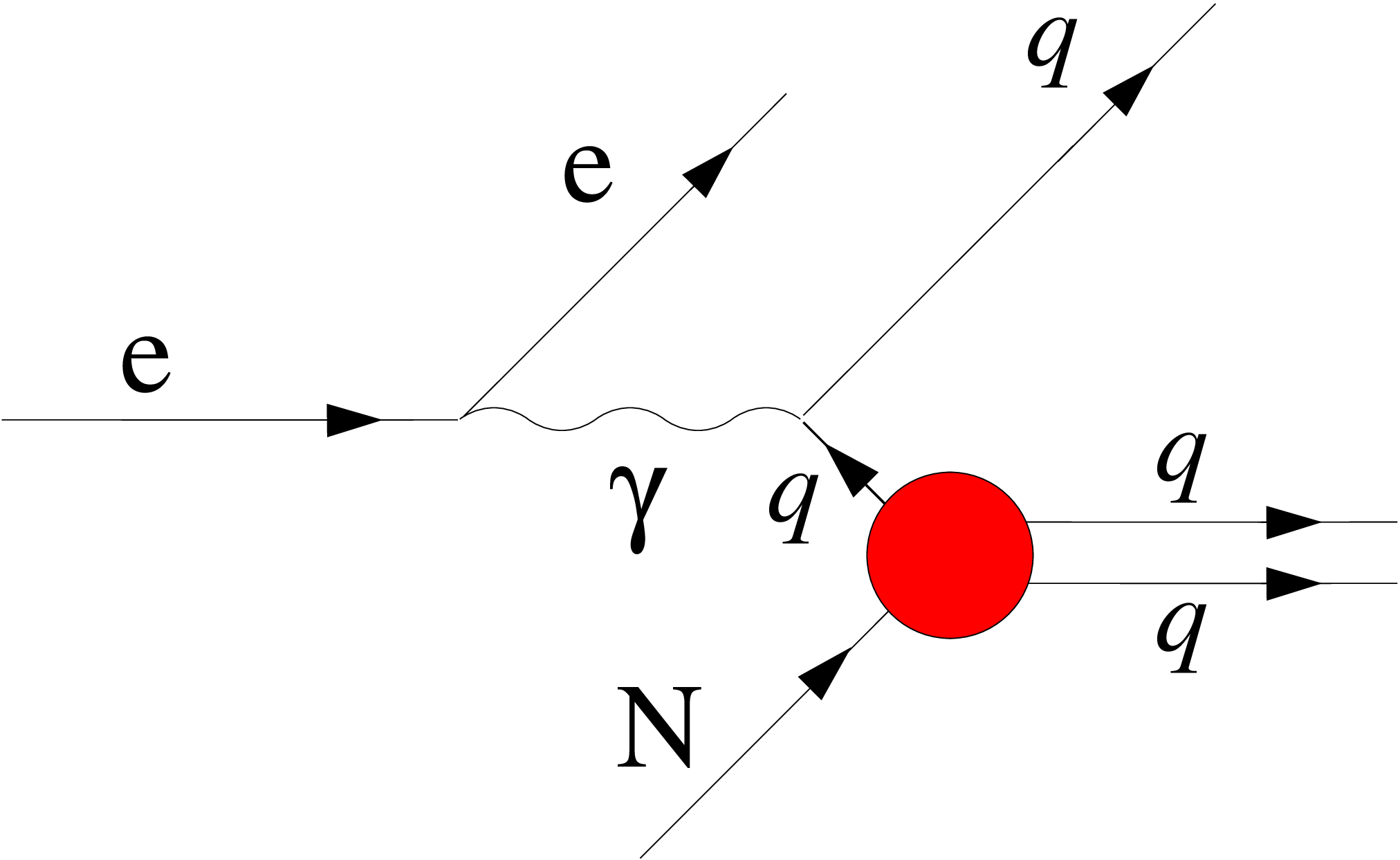}
\includegraphics[clip=true,width=0.3\columnwidth,angle=0.]
{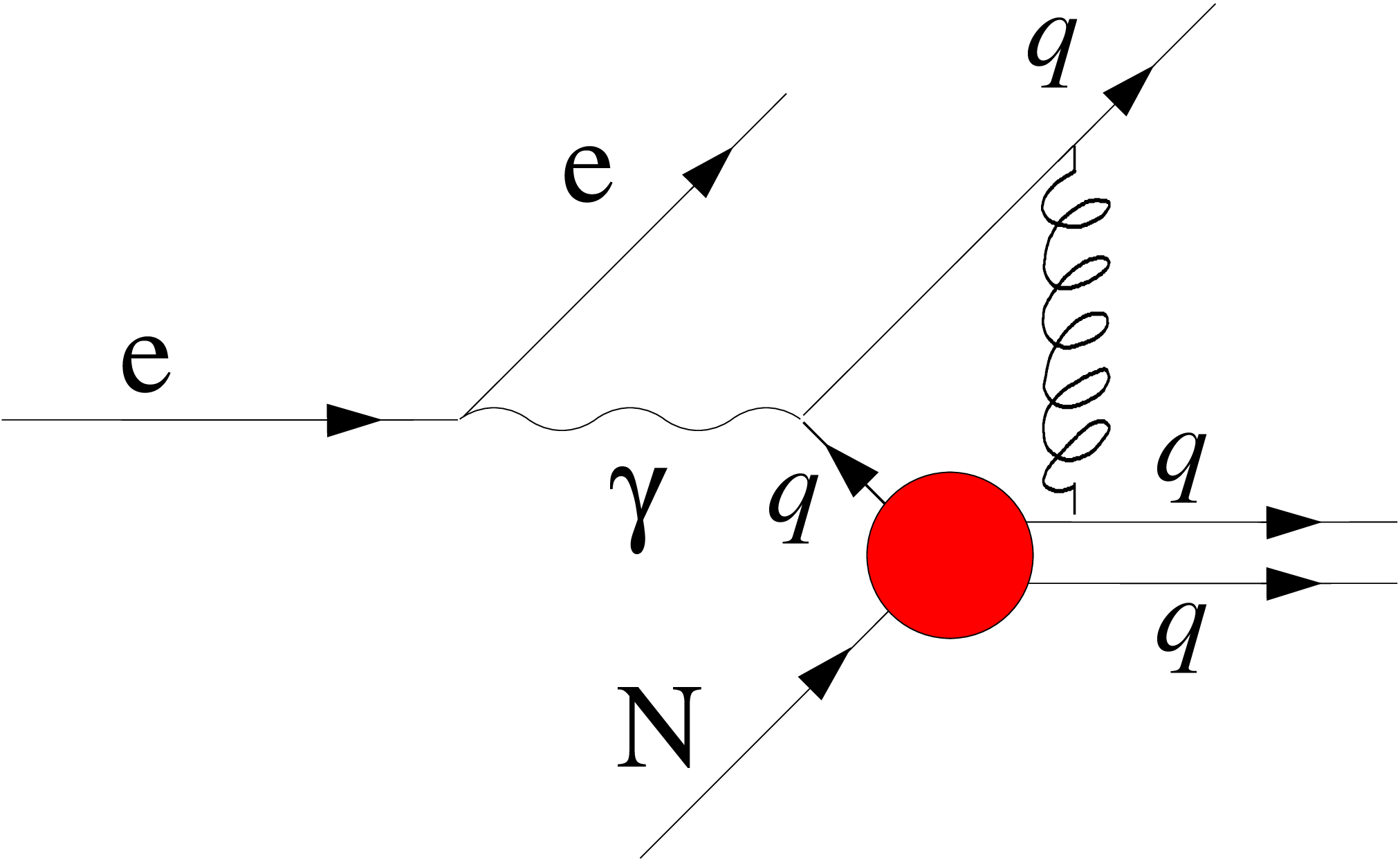}
\includegraphics[clip=true,width=0.3\columnwidth,angle=0.]
{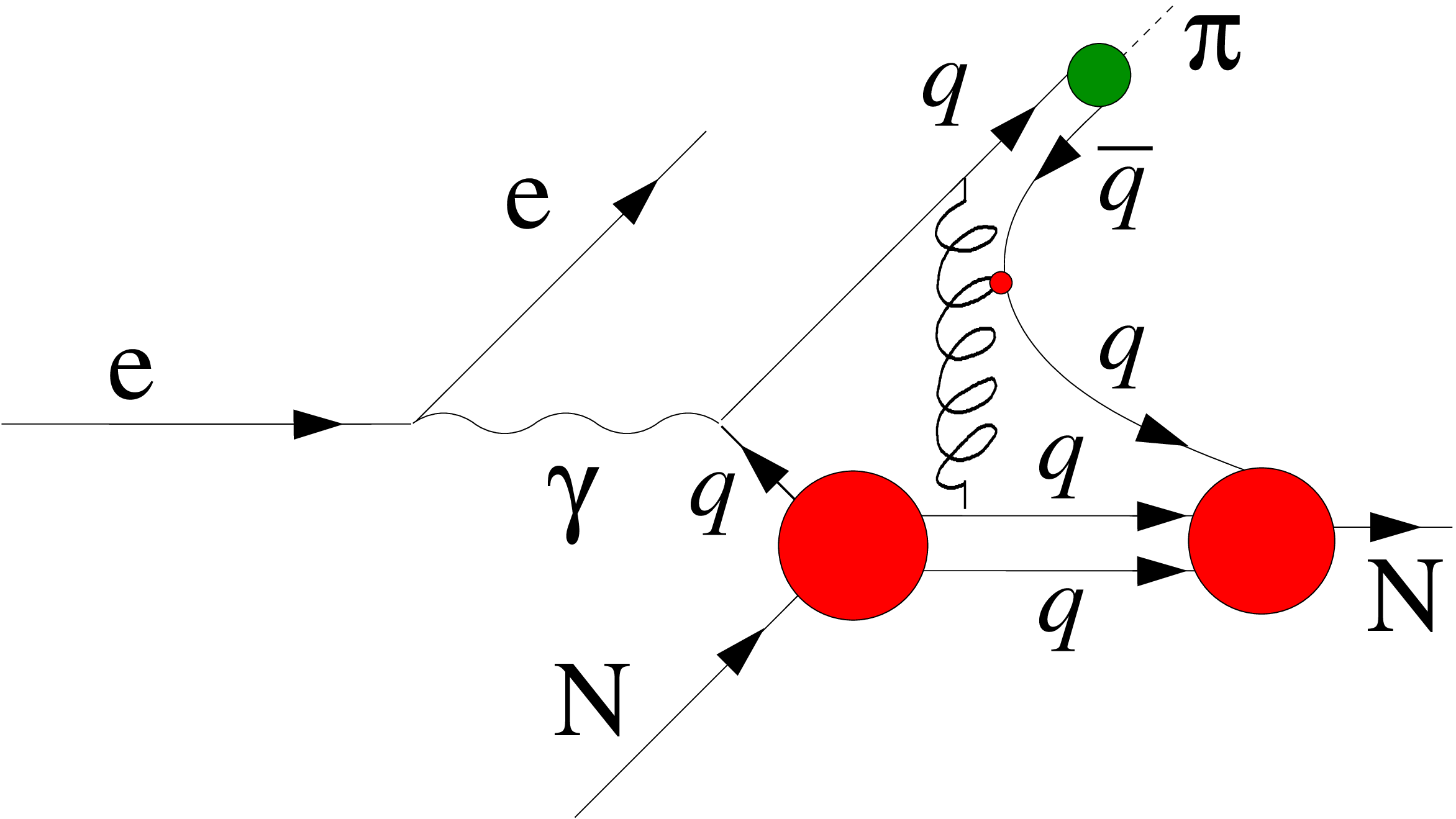}
\caption{\label{Figure2}
\small A schematic representation of the partonic DIS part of the $\pi^+$--
electroproduction mechanism. The wavy line represents a color string.
See text for the details.
\vspace{-0.8cm}
}
\end{center}
\end{figure}

\begin{figure}[t]
\begin{center}
\includegraphics[clip=true,width=0.65\columnwidth,angle=0.]{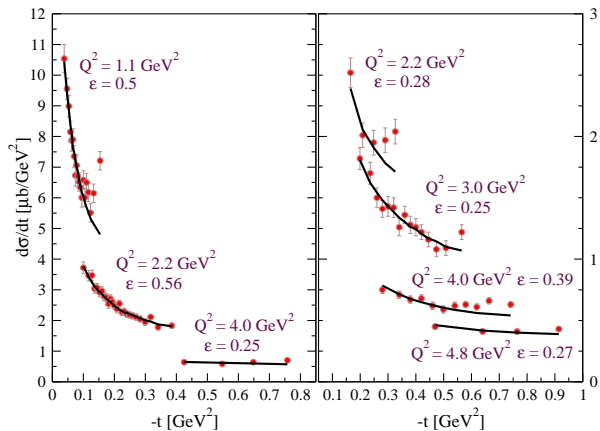}
\caption{\label{Figure4} \small
Differential cross section  
$d\sigma/dt = d\sigma_{\rm T}/dt + \varepsilon d\sigma_{\rm
  L}/dt$ of the reaction $p(\gamma^*,\pi^+)n$. 
The solid curves are the model predictions. The experimental data are 
from Ref.~\cite{:2009ub}.
\vspace{-0.5cm}
}
\end{center}
\end{figure}

For the description of the reaction $p(e,e'\pi^+)n$ in DIS 
a model for the hadronization process is needed. In the present 
description of the hadronization in DIS we rely on the Lund fragmentation
model
as depicted in Figure~\ref{Figure2}
where the leading order $\gamma^* q \to q$ DIS process followed by the 
fragmentation of an excited string into two particles ($\pi N$)
is shown. As a
realization of the Lund model we use the \Jetset{} implementation.
This description which connects the exclusive and
semi-inclusive DIS production in the limit $z\to 1$ resolves the longstanding puzzle of a
large theoretical underestimate of the observed transverse strength in
the reaction $p(e,e'\pi^+)n$~\cite{Kaskulov:2008xc}. For instance, in Fig.~\ref{Figure4} we confront the result of our calculations
(solid curves) with the JLAB data~\cite{:2009ub} 
for unseparated cross sections at average value of $W \simeq 2.2$~GeV. 
The data are very well described by the
present model in a measured range from $Q^2 \simeq 1$~GeV$^2$ 
up to $5$~GeV$^2$. Note that the high $Q^2$ dependence of the data essentially follows 
the $Q^2$ dependence of the deep inelastic structure functions in DIS. A
strong rise at low values of $Q^2$ is due to the $\pi$-Reggeon exchange. Its
$Q^2$ dependence is driven by the pion charge form factor. The
high $Q^2$ domain of the reaction is totally transverse. See
Ref.~\cite{Kaskulov:2008xc} for further details.

Why DIS pions matter for the CT effect. A necessary condition for the CT effect is the
propagation of a quark--gluon system, originating in the hard partonic
interaction, through the nuclear medium and its subsequent interactions with
surrounding nucleons. In the present model the hard DIS part of the 
primary high energy electromagnetic
interaction is determined by the Lund model which means that the final state
consists of an excited string (see Fig.~\ref{Figure2}). This string then
fragments into hadrons. In the exclusive reaction $(e,e'\pi^+)$
considered here all DIS pions are -- because of  their high energy $z \approx 1$
-- directly connected to the hard interaction point and have production
time $t_P = 0$. Thus, following the Lund model hadronization pattern the
(pre)hadronic interaction in Eq.~(\ref{effcs}) is effective only for the DIS events;  the
longitudinal cross section, which is dominated by the 
$\pi$-knockout from the preexisting meson cloud of a nucleon, is not affected by this (pre)hadronic
interaction. In the model of Ref.~\cite{Kaskulov:2008xc} 
only the DIS (transverse) part of the cross section is
responsible for the observed CT effect since this part is connected
with the 4D pattern of the string breaking dynamics which makes the
formation time of produced (pre)hadrons finite. Indeed, the transverse nature of the
CT effect at JLAB agrees remarkably well with data, see Fig.~\ref{Figure8}.

\begin{center}
\large{4. \textit{Deep exclusive $\pi$ electroproduction and nucleon resonances}}
\end{center}
In~\cite{Kaskulov:2008xc,Kaskulov:2009gp} the transverse cross
section $\sigma_{\rm T}$ was modeled using the string breaking mechanism in
DIS. However, the solution of the problem 
on the amplitude level is still missing.  Both the soft hadronic and hard 
partonic parts of the amplitude can in principle interfere making non-additive 
contributions to $\sigma_{\rm L}$ and to interference $\sigma_{\rm TT}$ and 
$\sigma_{\rm LT}$ cross sections. 
 Indeed, the data from JLAB 
demonstrate~\cite{Horn:2007ug,Blok:2008jy} that the magnitude and sign of the
interference cross sections are not compatible with the simple exchange of a 
pion Regge trajectory in the $t$-channel. Because, the contributions from exchange
of heavy mesons are small~\cite{Kaskulov:2008xc} this would suggest the
presence of a large transverse resonance or partonic  interfering background 
to the meson-pole contributions.

\begin{figure}[t]
\begin{center}
\includegraphics[clip=true,width=0.65\columnwidth,angle=0.]{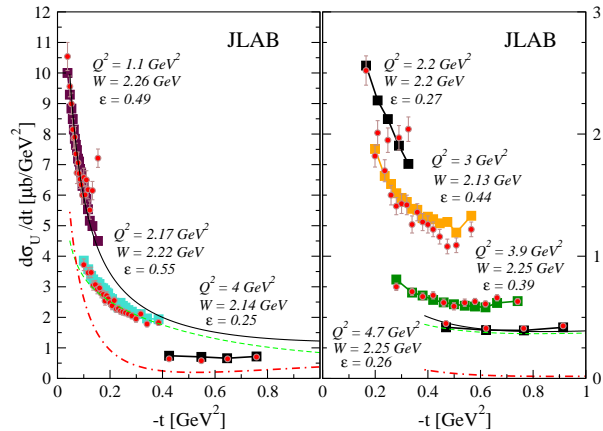}
\caption{\label{Dutta1} \small The differential cross sections
$d\sigma_{\rm U}/dt = d\sigma_{\rm T}/dt + \varepsilon d\sigma_{\rm
L}/dt$ in exclusive reaction $p(\gamma^*,\pi^+)n$
at JLAB~\cite{:2009ub}.
The square symbols connected by solid lines describe the
model results. The discontinuities in the curves result from
the different values of $(Q^2,W,\varepsilon)$ for the various $-t$ bins. 
The dash-dotted and dashed curves describe the contributions of the longitudinal
$\varepsilon d\sigma_{\rm L}$ and transverse $d\sigma_{\rm T}$
cross sections, respectively, to the total unseparated cross sections (solid curves)
for the the lowest and highest average values of $Q^2=1.1$~GeV$^2$ and $Q^2=4.7$~GeV$^2$.}
\vspace{-0.5cm}
\end{center}
\end{figure}

In Ref.~\cite{KM} we attempted a phenomenological approach to model 
the presence of nucleon resonances beyond the $t$-channel meson-exchange amplitudes. 
We modeled the contribution of nucleon resonances using a local Bloom-Gilman
connection between the exclusive and inclusive reactions.

\begin{figure}[t]
\begin{center}
\includegraphics[clip=true,width=1\columnwidth,angle=0.]
{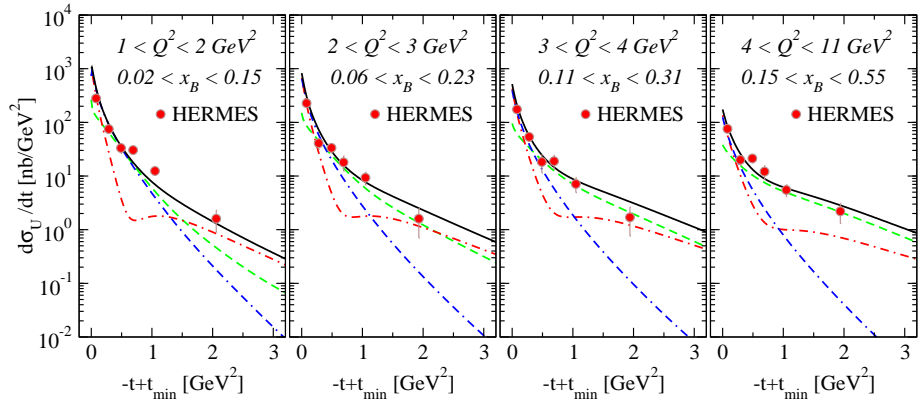}
\caption{\label{EffHermes} \small $-t+t_{min}$ dependence of the 
differential cross section $d\sigma_U/dt = d\sigma_{\rm T}/dt + \epsilon d\sigma_{\rm
L}/dt$ in exclusive reaction $p(\gamma^*,\pi^+)n$ at HERMES. 
The experimental data are from Ref.~\cite{:2007an}. 
The calculations are performed for the average values of
$(Q^2,x_{\rm B})$ in a given $Q^2$ and Bjorken $x_{\rm B}$ bin.
The solid curves are the full model results.
The dash-dotted curves correspond to the longitudinal $\epsilon d\sigma_{\rm L}/dt$
and the dashed curves to the transverse $d\sigma_{\rm T}/dt$ components of the
cross section.
The dash-dash-dotted curves describe the results without the
resonance/partonic effects.}
\vspace{-0.8cm}
\end{center}
\end{figure}

In Fig.~\ref{Dutta1} we show the results of Ref.~\cite{KM} where the
excitation of nucleon resonances is taken into account. Also compare these
results with the ones obtained in the previouse section, see
Fig.~\ref{Figure4}, when using the concept of DIS pions.
The square symbols connected by solid lines describe the
model results. 
The data are very well reproduced by the
present model in the measured $Q^2$ range from $Q^2 \simeq 1$~GeV$^2$
up to $5$~GeV$^2$.  
In Fig.~\ref{Dutta1} 
we also show the contributions of the longitudinal
$\varepsilon d\sigma_{\rm L}$ (dash-dotted curves) and transverse $d\sigma_{\rm T}$
(dashed curves) cross sections to the total unseparated cross sections (solid curves)
for the lowest and highest average values of $Q^2=1.1$~GeV$^2$ and $Q^2=4.7$~GeV$^2$.
The cross sections at high values of $Q^2$ are flat and
totally transverse. At forward angles a strong peaking of the cross
section at $Q^2=1.1$~GeV$^2$ comes from the large
longitudinal component in this case. The off-forward region is
transverse. This behavior agrees with the results from~\cite{Kaskulov:2008xc}.

\begin{figure}[t]
\begin{center}
\includegraphics[clip=true,width=0.6\columnwidth,angle=0.]{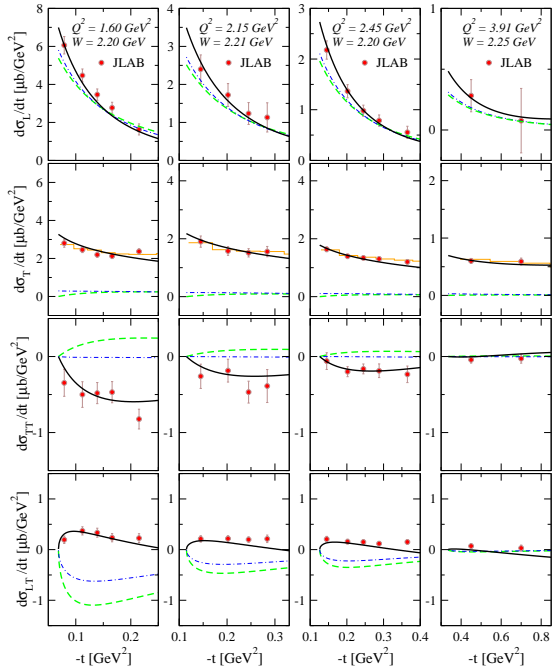}
\caption{\label{Horn1} 
\small $-t$ dependence of \textsc{l/t} partial transverse 
$d\sigma_{\rm T}/dt$, longitudinal $d\sigma_{\rm L}/dt$ and interference 
$d\sigma_{\rm TT}/dt$ and $d\sigma_{\rm LT}/dt$ differential cross sections 
in exclusive reaction $p(\gamma^*,\pi^+)n$. The experimental data are from
the $F\pi$-2~\cite{Horn:2006tm} and $\pi$-CT~\cite{Horn:2007ug}
experiments at JLAB. The numbers displayed in the plots are the average $(Q^2,W)$ values. 
The dashed curves correspond to the exchange of the $\pi$-Regge trajectory alone.
The dash-dotted curves are obtained with the on-mass-shell form
factors in the nucleon-pole contribution and exchange of the
$\rho(770)/a_2(1320)$-trajectory. The solid curves describe the model results with the
resonance contributions. The data points in each $(Q^2,W)$ bin correspond 
to slightly different values of $Q^2$ and $W$ for the various $-t$ bins.
The calculations are performed for values of $Q^2$ and $W$ corresponding to 
the first $-t$ bin. The histograms for $d\sigma_{\rm T}/dt$ are the results 
from~\cite{Kaskulov:2008xc}. 
\vspace{-0.5cm}
}
\end{center}
\end{figure}

The same behavior is observed in the DIS
regime at HERMES~\cite{:2007an} where the value of $W$ is
higher. At HERMES, because of the Regge shrinkage
of the $\pi$-reggeon exchange and smaller transverse component, the forward peak
just has a steeper $-t$-dependence~\cite{Kaskulov:2009gp}.
In Fig.~\ref{EffHermes} we show our results
for the $-t+t_{min}$ dependence of the 
differential cross section $d\sigma_U/dt = d\sigma_{\rm T}/dt + \epsilon d\sigma_{\rm
L}/dt$ in exclusive reaction $p(\gamma^*,\pi^+)n$ at HERMES.

In Figure~\ref{Horn1} we show our results for separated differential cross
sections in the $p(\gamma^*,\pi^+)n$ reaction together with the high-$Q^2$ data 
from~\cite{Horn:2006tm,Horn:2007ug}. The notations for the curves 
are described in the caption to the figure. The histograms in Fig.~\ref{Horn1} for $d\sigma_{\rm T}/dt$ are 
the results of Ref.~\cite{Kaskulov:2008xc}. Our present treatment of 
resonance contributions produces a result which is very close to that 
obtained in our previous work~\cite{Kaskulov:2008xc}. However,
the present approach goes beyond the two-component  hadron-parton model of 
Ref.~\cite{Kaskulov:2008xc} and allows to study the interference and 
non-$\pi$-pole background effects on the amplitude level.

\begin{figure}[t]
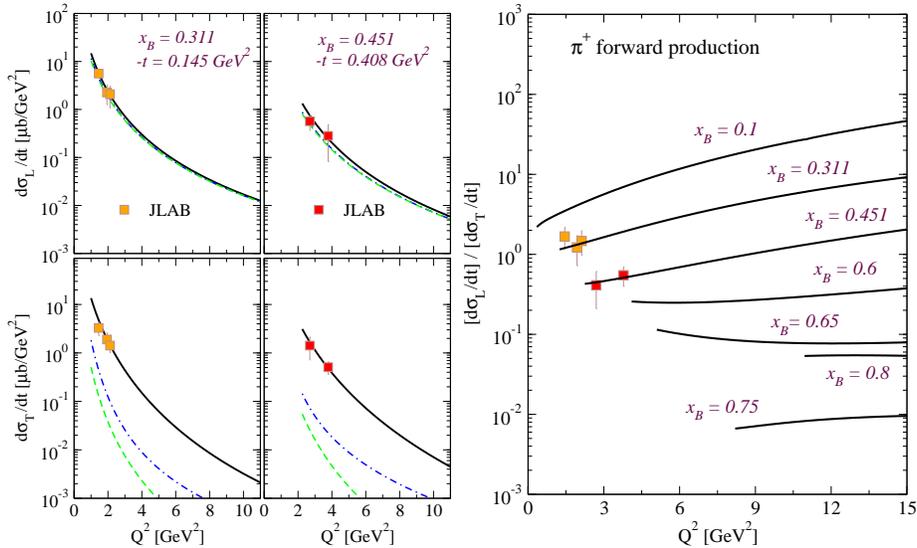

\includegraphics[clip=true,width=0.49\columnwidth,angle=0.]
{eFig17.eps}
\includegraphics[clip=true,width=0.5\columnwidth,angle=0.]
{eFig18.eps}
\caption{\label{EffHornScalingQ2} 
\small Left panel: $Q^2$ dependence of the longitudinal $d\sigma_{\rm
  L}/dt$ (top panels) and transverse $d\sigma_{\rm T}/dt$ (bottom panels) 
cross sections in $p(\gamma^*,\pi^+)n$ reaction at fixed values of
$-t$ and Bjorken $x_{\rm B}$. The solid curves are the model predictions for
the scaling curves. The dashed curves correspond to the contribution 
of the $\pi$-reggeon exchange alone. The dash-dotted curves are the model
results without the contributions of resonances.
The experimental data are from Ref.~\cite{Horn:2007ug}. Right panel: 
The $Q^2$ dependence of the ratio of longitudinal $d\sigma_{\rm L}/dt$ to transverse $d\sigma_{\rm
  T}/dt$ differential cross sections in the forward $\pi^+$ production. 
The different curves correspond to different values of Bjorken scaling
variable $x_{\rm B}$.
\vspace{-0.4cm}
}
\end{figure}

\begin{center}
\large{5. \textit{$Q^2$ dependence of the cross sections}}
\end{center}

It has been proposed that the $Q^2$ dependence of $\textsc{l/t}$ separated exclusive
$p(\gamma^*,\pi^+)n$ cross sections may provide a test of the factorization 
theorem~\cite{Collins:1996fb} in the separation of long-distance and 
short-distance physics and the extraction of GPD.  The leading twist GPD 
scenario predicts for $\sigma_{\rm L}\sim 1/Q^6$
and $\sigma_{\rm T}\sim 1/Q^8$. An observation of the $Q^2$ power law scaling
is considered as a model independent test of QCD factorization.

The $Q^2$ behavior of cross sections in exclusive reaction $p(\gamma^*,\pi^+)n$ 
has been studied at JLAB in Ref.~\cite{Horn:2007ug}. 
It was shown that while the scaling laws are
reasonably consistent with the $Q^2$ dependence of the longitudinal $\sigma_{\rm L}$ data,
they fail to describe the $Q^2$ dependence of the transverse $\sigma_{\rm T}$ data.
The $Q^2$ dependence of the $p(\gamma^*,\pi^+)n$ cross section in DIS has been
also studied at HERMES~\cite{:2007an}. It was found that the $Q^2$
dependence
of the data is in general well described by the calculations from GPD models
which include the power corrections,
see Ref.~\cite{:2007an} and references therein. However, the magnitude of the theoretical cross
section is underestimated. 
In the following we check this predicted $\sigma_{\rm L} / \sigma_{\rm T}
\sim Q^2$ scaling within our model calculations.

In Fig.~\ref{EffHornScalingQ2} (left panel) we show our results for the $Q^2$ 
dependence of $p(\gamma^*,\pi^+)n$ reaction cross sections 
$d\sigma_{\rm L}/dt$ and $d\sigma_{\rm T}/dt$ at fixed $-t$ and
Bjorken variable $x_B$. The experimental data are from Ref.~\cite{Horn:2007ug}
and correspond to the forward $\pi^+$ production. The solid curves 
are the model predictions and describe the available data
very well. The dashed curves describe the contribution of the
$\pi$-reggeon exchange to the $Q^2$ scaling curves only. The dash-dotted 
curves are the model results without the resonance contributions. 
The latter effect is again large in the transverse cross section and 
gives only small correction to the longitudinal cross section 
$d\sigma_{\rm L}/dt$. The $Q^2$ dependence of $d\sigma_{\rm L}/dt$ is 
essentially driven by the pion form factor.

\begin{figure}
\includegraphics[clip=true,width=1\columnwidth,angle=0.]
{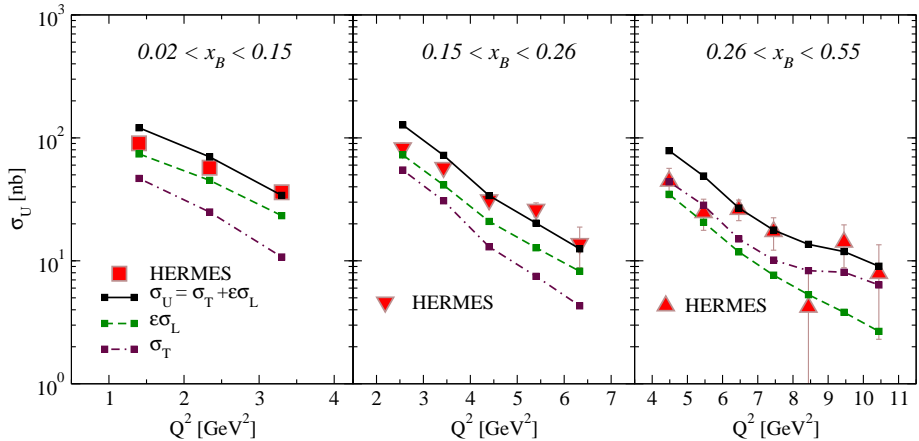}
\caption{\label{HermesQ2dep}  \small 
$Q^2$ dependence of the integrated cross sections 
$\sigma_{\rm U}=\sigma_{\rm T}+\epsilon \sigma_{\rm L}$
in exclusive reaction $p(\gamma^*,\pi^+)n$ at HERMES. 
The different panels correspond to different $x_B$ bins. 
The solid curves are the model results. 
The dashed and dash-dotted curves correspond to the longitudinal $\epsilon \sigma_{\rm L}$
and transverse $\sigma_{\rm T}$ components
of the cross section, respectively. The experimental data are from
Ref.~\cite{:2007an}} 
\vspace{-0.4cm}
\end{figure}

The $Q^2$ dependence of the ratio of longitudinal $d\sigma_{\rm L}/dt$ to 
transverse $d\sigma_{\rm  T}/dt$ differential cross sections for the 
forward $\pi^+$ production is shown in Fig.~\ref{EffHornScalingQ2} (right panel). 
The different curves correspond to different values of $x_{\rm B}$. 
All the curves start at the value of $W \simeq 1.9$~GeV. For small and 
intermediate values of $x_{\rm B}$ the model results show an increase of 
the ratio $d\sigma_{\rm L}/d\sigma_{\rm T}$ as a function of $Q^2$. Only at 
small values of Bjorken $x_{\rm B}$ the ratio $d\sigma_{\rm L}/d\sigma_{\rm
  T}$ is qualitatively in agreement with the predicted $\sim Q^2$ behavior.  
In the valence quark region above $x_{\rm B} \simeq 0.6$ the cross
section ratio scales and is actually independent of the value of $Q^2$. In
this region the transverse component  $\sigma_{\rm T}$ dominates the $\pi^+$ 
electroproduction cross section. In the experimental determination of the pion
transition form factor from forward $\sigma_{\rm L}$ data one can, therefore,
better isolate the longitudinal response by minimizing the Bjorken $x_{\rm B}$.

The physics content of the HERMES
deep exclusive $p(\gamma^*,\pi^+)n$ data is essentially the same as at JLAB.
Fig.~\ref{HermesQ2dep} shows the $Q^2$ dependence of 
the measured cross sections in DIS for different 
$x_{\rm B}$ bins~\cite{:2007an}. 
These are the same data sets from HERMES (see previous section) integrated over $-t$. 
The $Q^2$ dependence of the 
experimental data is well described by the calculations (solid curves) from 
the present model. The dashed and dash-dotted curve describe the longitudinal
$\varepsilon \sigma_{\rm L}$ and the transverse $\sigma_{\rm T}$ components, respectively.

In the region of small Bjorken $x_{\rm B}$, see left panel in
Fig.~\ref{HermesQ2dep}, the integrated longitudinal component dominates over
the transverse cross section. With increasing $x_{\rm B}$ the strength of the
transverse component is increasing and for values of $x_{B}$ in the third bin
the transverse cross section $\sigma_{\rm T}$ becomes the dominant part of the 
exclusive cross section. An increase of the relative contribution of 
$\sigma_{\rm T}$ as a function of $x_{\rm B}$ can be clearly seen in the 
right panel of Figure~\ref{HermesQ2dep} where $0.26<x_{\rm B}<0.55$.
There the first $Q^2$ bin corresponds to the average value of $x_{\rm B}=0.29$ and
the last $Q^2$ bin to the average value of $x_{\rm B}=0.44$.

\begin{center}
\large{6. \textit{Beam single spin asymmetry (SSA)}}
\end{center}
We further consider the electroproduction reaction
\begin{equation}
\vec{e} + N \to e' + \pi + N
\end{equation}
and now assume that the target nucleon is unpolarized, whereas we allow
arbitrary polarization for the incoming electron.
With a polarized beam 
$\vec{e}$ and with an unpolarized target there is an additional 
component $\sigma_{\rm LT'}$~\cite{KM} in the $(e,e'\pi)$ cross section which is
proportional to the imaginary part of an interference between the \textsc{l/t}
photons and therefore sensitive to the relative phases of amplitudes.  

Using the polarized electron beam, the longitudinal beam single-spin asymmetry (SSA) in $N(\vec{e},e'\pi)N'$ scattering is defined
so that
\begin{equation}
\label{BSSA}
A_{\rm LU}(\phi) \equiv
\frac{d\sigma^{\rightarrow}(\phi)-d\sigma^{\leftarrow}(\phi)}{d\sigma^{\rightarrow}(\phi)+d\sigma^{\leftarrow}(\phi)},
\end{equation}
where $d\sigma^{\rightarrow}$ refers to positive helicity $h=+1$ of the incoming
electron. The azimuthal moment associated with the beam SSA is
given by
\begin{equation}
\label{BeamSSAmoment}
A^{\sin(\phi)}_{\rm LU} = \frac{\sqrt{2\varepsilon(1-\varepsilon)}d\sigma_{\rm
    LT'}}{d\sigma_{\rm T} + \varepsilon d\sigma_{\rm L}}.
\end{equation}

In general, a nonzero $\sigma_{\rm LT'}$ or the corresponding 
beam SSA $A_{\rm LU}(\phi)$, Eq.~(\ref{BSSA}), demands 
interference between single helicity flip and nonflip or double helicity flip 
amplitudes. In Regge models the asymmetry may result from 
Regge cut corrections to single reggeon exchange. 
This way the amplitudes in the product acquire different phases and therefore 
relative imaginary parts. A nonzero beam SSA  can be also generated by the 
interference pattern of amplitudes where particles with opposite
parities are exchanged.

\begin{figure}[t]
\begin{center}
\includegraphics[clip=true,width=0.55 \columnwidth,angle=0.]{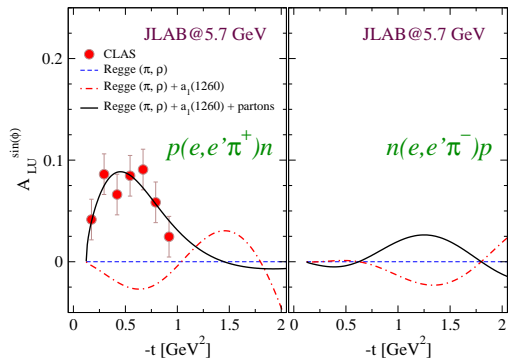}
\caption{\label{BSAAvakian}
\small  Left panel: The beam spin azimuthal moment
$A^{\sin(\phi)}_{\rm LU}$ in exclusive
reaction $p(\gamma^*,\pi^+)n$ as a function of $-t$. 
The CLAS/JLAB data are from~\cite{Avakian:2004dt}.  
The dashed curves describe 
the results (the asymmetry is zero) without the resonance contributions 
and neglecting the exchange of unnatural parity  $a_1(1260)$ Regge trajectory.  
The dash-dotted curves correspond
to the addition of the axial-vector $a_1(1260)$-reggeon exchange. 
The solid curves are the model results and account for the
resonance/partonic effects.
Right panel:
The beam spin azimuthal moment $A^{\sin(\phi)}_{\rm LU}$ in exclusive reaction
$n(\gamma^*,\pi^-)p$.
\vspace{-0.5cm}
}
\end{center}
\end{figure}

In the left panel of Fig.~(\ref{BSAAvakian}) we plot the CLAS 
data~\cite{Avakian:2004dt} for the azimuthal moment $A^{\sin(\phi)}_{\rm LU}$ 
associated with the beam SSA, Eq.~(\ref{BeamSSAmoment}), in the reaction 
$p(\vec{e},e'\pi^+)n$.  These data have been collected in hard scattering
kinematics $E_e=5.77$~GeV, $W>2$~GeV and $Q^2>1.5$~GeV$^2$. 
The experiment shows a sizable and positive beam SSA. 

In the left and right panels of Fig.~(\ref{BSAAvakian}) 
we present our results for the azimuthal moments $A^{\sin(\phi)}_{\rm LU}$ 
in the reactions $p(\vec{e},e'\pi^+)n$ and $n(\vec{e},e'\pi^-)p$,
respectively. 

\begin{figure}[h]
\begin{center}
\includegraphics[clip=true,width=0.6\columnwidth,angle=0.]
{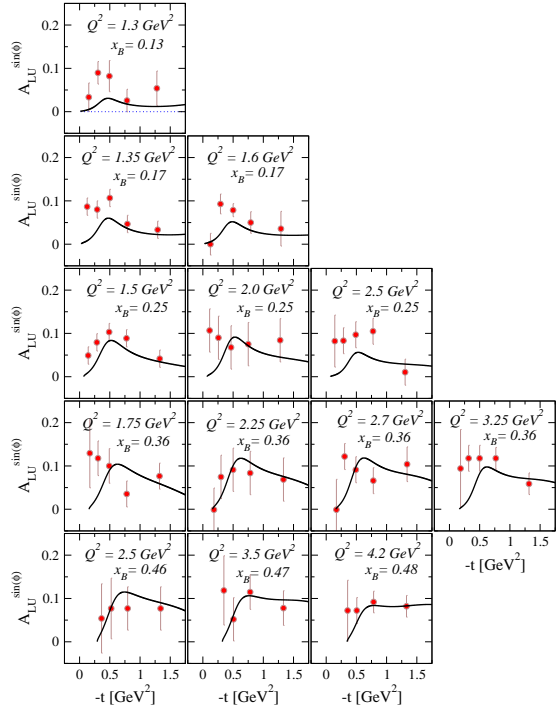}
\caption{\label{BeamSSApi0} 
\small The beam spin azimuthal moment
$A^{\sin(\phi)}_{\rm LU}$ in exclusive
reaction $p(\gamma^*,\pi^0)p$ as a function of $-t$ for different $(Q^2,x_{\rm
B})$ bins.
The solid curves are the model results and account for the
residual effect of nucleon resonances. The experimental data are from~\cite{DeMasi:2007id}.
}
\vspace{-0.5cm}
\end{center}
\end{figure}

At first, we consider $A^{\sin(\phi)}_{\rm LU}$ generated by the exchange of 
Regge trajectories. In Fig.~(\ref{BSAAvakian}) the dashed curves describe  
the model results without the effects of resonances and neglecting the exchange of the
axial-vector $a_1(1260)$ Regge trajectory. This model results in a zero
$A^{\sin(\phi)}_{\rm LU}$ and therefore a zero beam SSA. The addition of the unnatural
parity $a_1(1260)$-exchange generates by the interference with the natural parity 
$\rho(770)$ exchange a sizable $A^{\sin(\phi)}_{\rm LU}$ in both channels. This result
corresponds to the dash-dotted curves in Fig.~(\ref{BSAAvakian}). 
In the rest of unpolarized observables discussed above the effect of the axial-vector 
$a_1(1260)$ is small. However, as one can see, the contribution of $a_1(1260)$ 
is important in the polarization observables. For instance, a strong
interference pattern of the $a_1(1260)$-reggeon exchange makes the
polarization  observables, like the beam SSA,  very sensitive to the different 
scenarios describing the structure and behavior 
of $a_1(1260)$ in high-$Q^2$ processes.
In the last step we account for the resonance contributions. The latter
strongly influence the asymmetry parameter
$A^{\sin(\phi)}_{\rm LU}$. The model results (solid curves) are in agreement 
with the positive $A^{\sin(\phi)}_{\rm LU}$ in the
$\pi^+$ channel and predict much smaller $A^{\sin(\phi)}_{\rm LU}$ in the $\pi^-$
channel.  A sizable and positive $A^{\sin(\phi)}_{\rm LU}$ has been also observed at HERMES
in $\pi^+$ SIDIS close to the exclusive limit $z\to
1$~\cite{Airapetian:2006rx}.

As in deep exclusive $\pi^+$ electroproduction a sizeable and positive beam SSA   
has recently been  measured at CLAS/JLAB also in the exclusive reaction $p(\vec{e},e'\pi^0)p$~\cite{DeMasi:2007id}. 
It was shown that the simple Regge model used in~\cite{DeMasi:2007id} fails 
to explain the measured kinematic $(\sqrt{s},Q^2)$  dependencies. 
We have, therefore, extended our calculation of Ref.~\cite{KM} to the neutral pion channel. In the
Regge exchange contributions the vector $\omega(782)$ and axial-vector
$b_1(1235)$ and $h_1(1170)$ trajectories are taken into account. We find that at
high values of $Q^2$ the dominant contribution to the beam SSA again comes from
the residual excitation of nucleon resonances. Our results are shown in Fig.~\ref{BeamSSApi0} and
describe the JLAB data very well.

\begin{center}
\large{7. \textit{Summary}}
\end{center}
In the first part of this work we presented a calculation of the nuclear 
transparency of pions in the reaction $A(e,e'\pi^+)A^*$ off nuclei. The
microscopic input for the primary interaction of the virtual photon with 
the nucleon describes both the transverse and the longitudinal cross sections.
The coupled--channel BUU transport model has been used to describe the FSI of
hadrons in the nuclear medium. The formation times of (pre)hadrons follow the 
time--dependent hadronization pattern of hard DIS processes.
Our results are consistent with the JLAB data and show that a detailed
understanding of the primary $\gamma^*N$ interaction is essential for a
quantitative understanding (and proof) of CT.

In the second part of the work we described different approaches to exclusive
production of pions off nucleons. We have proposed a
resolution of the $\sigma_{\rm T}$ problem in
the 
reaction $p(e,e'\pi^+)n$ above the resonance region.
It is based on the concept of DIS pions and combines the meson-exchange
currents and DIS of virtual photons off partons.  
Following a two-component hadron-parton picture of Refs.~\cite{Kaskulov:2008xc,Kaskulov:2009gp}
we further developed a  model which combines a Regge pole approach with residual effects of nucleon
resonances. The contribution of nucleon
resonances has been assumed to be dual to direct partonic
interaction and therefore describes the hard part of the model cross
sections. The resonance/partonic effects are taken into account using a  
Bloom-Gilman connection between the exclusive hadronic form factors
and inclusive deep inelastic structure functions. The model results agree well
with the experimental data measured at JLAB and DESY. 

\begin{center}
\textit{Acknowledgements}
\end{center}
This work was supported by DFG through TR16 and by BMBF.
\noindent 

\end{document}